\documentclass[a4paper]{jpconf}
\usepackage{graphicx}
\begin{document}
\title{The ANAIS-112 experiment at the Canfranc Underground Laboratory}

\author{J Amar\'e, S Cebri\'an\footnote{Attending speaker.}, I Coarasa, C Cuesta\footnote{Present Address: Centro de Investigaciones Energ\'eticas, Medioambientales y
Tecnol\'ogicas, CIEMAT, 28040, Madrid, Spain.}, E Garc\'ia, M
Mart\'inez\footnote{Present Address: Universit\`a di Roma La
Sapienza, Piazzale Aldo Moro 5, 00185 Rome, Italy.}, M A Oliv\'an, Y
Ortigoza, A Ortiz de Sol\'orzano, J Puimed\'on, A Salinas, M L
Sarsa, P Villar and J A Villar\footnote{Deceased.}}

\address{Laboratorio de F\'isica Nuclear y Astropart\'iculas, Universidad de Zaragoza, Calle Pedro Cerbuna, 12, 50009 Zaragoza, Spain \\
Laboratorio Subterr\'aneo de Canfranc, Paseo de los Ayerbe s/n, 22880 Canfranc Estaci\'on, Huesca, Spain}

\ead{scebrian@unizar.es}

\begin{abstract}
The ANAIS experiment aims at the confirmation of the DAMA/LIBRA
signal at the Canfranc Underground Laboratory (LSC). Several 12.5~kg
NaI(Tl) modules produced by Alpha Spectra Inc. have been operated
there during the last years in various set-ups; an outstanding light
collection at the level of 15~photoelectrons per keV, which allows
triggering at 1~keV of visible energy, has been measured for all of
them and a complete characterization of their background has been
achieved.
In the first months of 2017, the full ANAIS-112 set-up consisting of
nine Alpha Spectra detectors with a total mass of 112.5~kg was
commissioned at LSC and the first dark matter run started in August,
2017. Here, the latest results on the detectors performance and
measured background from the commissioning run will be presented and
the sensitivity prospects of the ANAIS-112 experiment will be
discussed.
\end{abstract}

\section{Introduction}

The ANAIS (Annual modulation with NaI(Tl) Scintillators) experiment
intends to confirm the DAMA/LIBRA modulation signal
\cite{bernabei13} using the same target and technique in a different
environment, at the Canfranc Underground Laboratory (LSC,
Laboratorio Subterr\'aneo de Canfranc) in Spain. This goal imposes
strong experimental requirements: an energy threshold at or below
2~keV$_{ee}$\footnote{Electron equivalent energy.}, background as
low as possible below~10 keV$_{ee}$ 
and very stable operation conditions.
Since the nineties, NaI(Tl) detectors from different suppliers have
been in operation in Canfranc~\cite{sarsa97,anaisepjc}; Alpha
Spectra Inc. (AS) detectors have shown best performance and
radiopurity (characterized in set-ups with two or three detectors,
referred as ANAIS-25~\cite{anaisnima} and
ANAIS-37~\cite{anaistaup15}, respectively) and then, they have been
selected to be used in ANAIS. In 2017, ANAIS-112 experiment has been
commissioned: the ANAIS-112 set-up consists of a 3$\times$3 matrix
of 12.5~kg NaI(Tl) modules and data taking is underway since August,
2017 for at least the next two years in the same conditions. A blind
annual modulation analysis is foreseen, as well as the ANAIS data
public release after their scientific exploitation. The set-up of
ANAIS-112 is described in section~\ref{exp}. Results concerning the
detector performance and background are presented in
sections~\ref{res} and \ref{bac}. Finally, the sensitivity expected
in the search for an annual modulation signal is discussed in
section~\ref{sen}.

\section{Experimental set-up}
\label{exp}

\begin{figure}
\begin{center}
\begin{minipage}{17pc} 
\includegraphics[width=\textwidth]{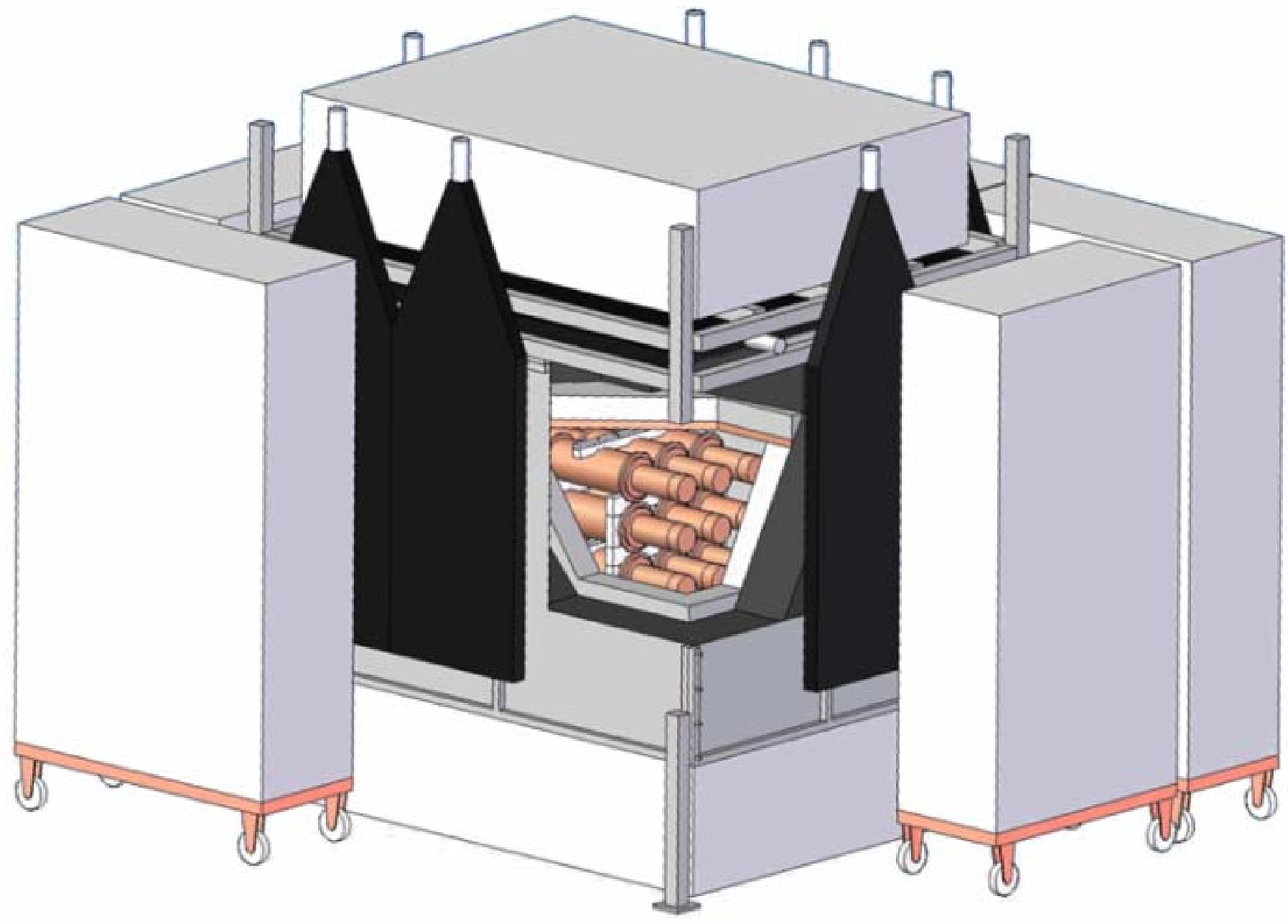}
\caption{\label{ANAISSetup} Design of the ANAIS-112 set-up mounted
at LSC (see text).}
\end{minipage}\hspace{1pc}%
\begin{minipage}{17pc} 
\includegraphics[width=\textwidth]{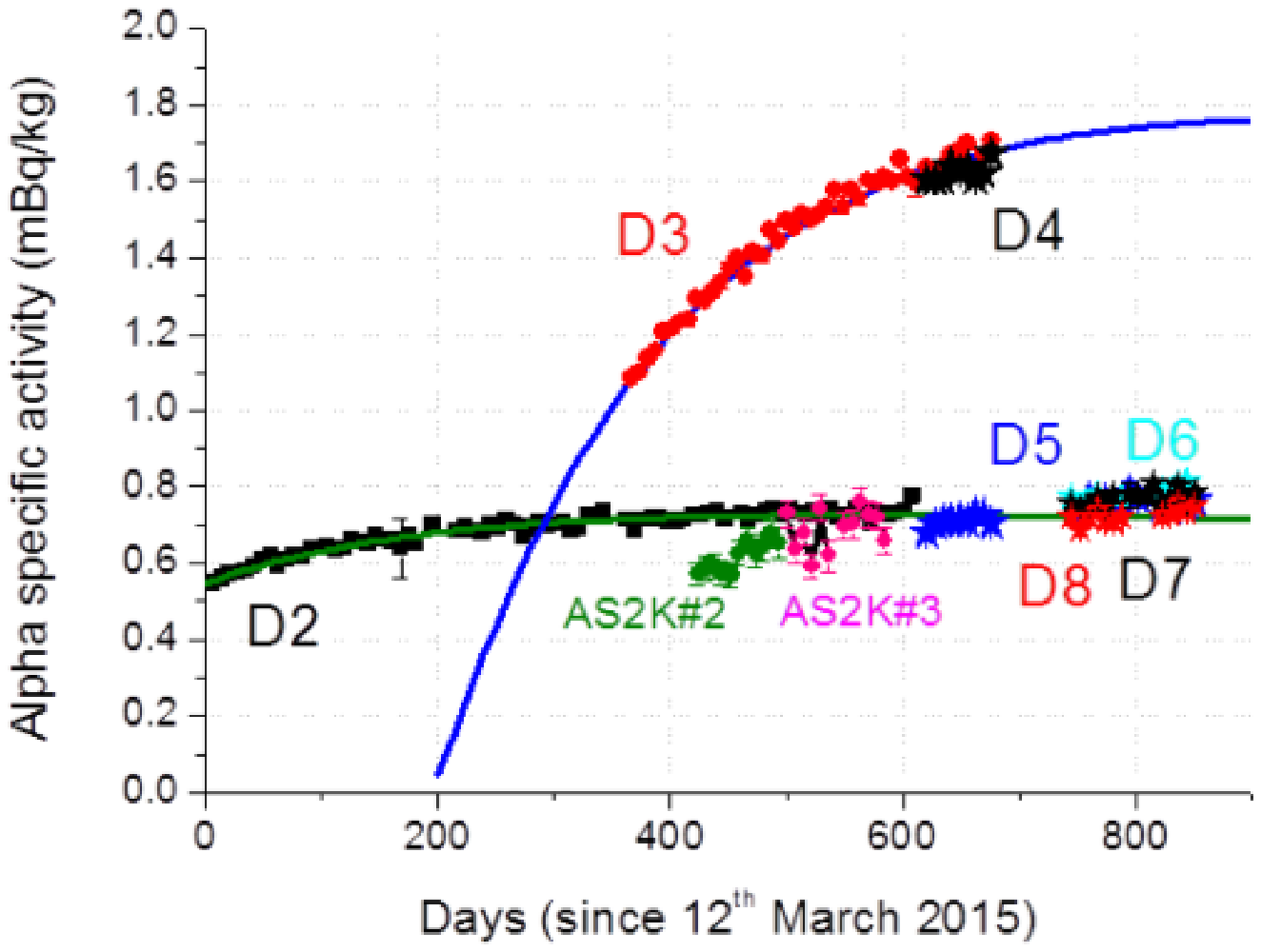}
\caption{\label{alphas} Evolution in time of the alpha specific
activity for different crystals as well as for smaller samples
(AS2K\#2-3) analyzed in the selection of material.}
\end{minipage}
\end{center}
\end{figure}

The nine modules used in ANAIS-112 were produced by AS in Colorado
and then shipped to Spain along several years, arriving at LSC the
first of them at the end of 2012 and the last by March, 2017 (see
table~\ref{detectors}). Each crystal is cylindrical (4.75'' diameter
and 11.75'' length), with a mass of 12.5~kg. NaI(Tl) crystals were
grown from selected ultrapure NaI powder and housed in OFE (Oxygen
Free Electronic) copper; the encapsulation has a mylar window
allowing low energy calibration. Two Hamamatsu R12669SEL2
photomultipliers (PMTs) were coupled through quartz windows to each
crystal at LSC clean room. All PMTs have been screened for
radiopurity using germanium detectors in Canfranc. The shielding for
the experiment consists of 10~cm of archaeological lead, 20~cm of
low activity lead, 40~cm of neutron moderator, an anti-radon box (to
be continuously flushed with radon-free air) and an active muon veto
system made up of plastic scintillators designed to cover top and
sides of the whole ANAIS set-up (see figure~\ref{ANAISSetup}).
The hut housing the experiment is at the hall B of LSC under
2450~m.w.e..

\begin{table}[h]
\caption{\label{detectors} Features of the nine ANAIS detectors
produced by Alpha Spectra: type of NaI(Tl) powder used, date of
arrival at LSC, total light collection measured in the ANAIS-112
set-up (except for D1 (*), from previous set-ups) and deduced
activity of $^{40}$K and $^{210}$Pb. Results for detectors D4-D8,
obtained from the commissioning run of ANAIS-112, are preliminary.}
\begin{center}
\begin{tabular}{llllll}
\br
Detector &  Quality powder & Arrival date & Light collection & $^{40}$K activity & $^{210}$Pb activity  \\
& & & (phe/keV) & (mBq/kg) & (mBq/kg) \\ \mr
D0 & $<$90 ppb K & December 2012 & 15.3$\pm$1.1 & 1.1 & 3.15 \\
D1 & $<$90 ppb K & December 2012 & 14.8$\pm$0.5 * & 1.4 & 3.15  \\
D2 & WIMPScint-II & March 2015 &  15.3$\pm$1.4 & 0.9& 0.70 \\
D3 & WIMPScint-III & March 2016 & 14.6$\pm$0.8 & 1.0&1.8 \\
D4  & WIMPScint-III & November 2016 & 14.0$\pm$0.8 & 1.0& 1.8\\
D5  & WIMPScint-III & November 2016 & 14.0$\pm$0.8 & 1.0& 0.75\\
D6 & WIMPScint-III & March 2017 & 12.6$\pm$0.8 & 1.1&0.76 \\
D7 & WIMPScint-III & March 2017 & 17.0$\pm$2.0& 1.0&0.75 \\
D8 & WIMPScint-III & March 2017 & 14.6$\pm$0.9& 0.6&0.72 \\
\mr
average & & &  & 1.0 & 1.5 \\
\br
\end{tabular}
\end{center}
\end{table}

DAQ hardware and software of ANAIS-112 were tested in previous ANAIS
set-ups. For each module, individual PMT charge output signals are
digitized and fully processed. Triggering is done by the coincidence
(logical AND) of the two PMT signals of any detector at
photoelectron level in a 200~ns window, enabling digitization and
conversion of the two signals. There is redundant energy conversion
by QDC modules and the building of the spectra is done off-line by
adding the signals from both PMTs. The muon detection system based
on plastic scintillators is fully implemented, allowing to tag
muon-related events and to monitor on-site the muon flux. The slow
control system is also operative, monitoring different parameters
like radon activity, humidity, pressure, several temperatures,
N$_{2}$ flux or PMT High Voltage. In addition, a blank module will
be set-up to monitor non-NaI(Tl) scintillation events and build a
``blank'' population for the study of annual modulation systematics.

\section{Detector performance}
\label{res}

The light output measured for all AS modules is at the level of
$\sim$15~phe/keV, which is a factor of two larger than that
determined for the best DAMA/LIBRA detectors \cite{bernabei2008}.
The fourth column of table~\ref{detectors} shows the preliminary
results for the total number of photoelectron per keV using
ANAIS-112 data, following the same method applied in ANAIS-25 and
ANAIS-37 set-ups, described in \cite{APlightyield}. The new estimate
is in very good agreement with the previous ones for D0-D5
detectors. This high light collection, possible thanks to the
excellent crystal quality and the use of high quantum efficiency
PMTs, has a direct impact in energy threshold.
Triggering below 1~keV$_{ee}$ is confirmed by the identification of
bulk $^{22}$Na and $^{40}$K events at 0.9 and 3.2~keV$_{ee}$,
respectively, thanks to coincidences with the corresponding high
energy photons following the electron capture decays to excited
levels.

Effective filtering protocols for rejecting non-bulk scintillation
events, similar to those described at \cite{anaisepjc} and optimized
for each detector, have been applied. Multiparametric cuts based on
the number of peaks in the pulses, the temporal parameters of the
pulses and the asymmetry in light sharing between PMTs are
considered. Acceptance efficiency curves from external calibration
data are obtained for each detector.


\section{Radiopurity and background}
\label{bac}

Detailed background models for the first modules operated in
ANAIS-25 and ANAIS-37 set-ups were developed \cite{anaisbkg}, based
on Geant4 Monte Carlo simulations and an accurate quantification of
background sources: the intrinsic crystal activity directly
assessed, the cosmogenic activity in crystals (precisely quantified
from ANAIS-25 data \cite{anaisjcap}) and the activity from external
components measured with HPGe detectors in Canfranc. At the region
of interest, crystal bulk contamination is the dominant background
source. Contributions from $^{40}$K and $^{22}$Na peaks and the
continua from $^{210}$Pb and the considered cosmogenic $^{3}$H are
the most relevant ones.

The ANAIS-112 data taken up to July, 2017 in the commissioning run
have been analyzed to make a first quantification of the relevant
background sources.  The activity of $^{40}$K and $^{210}$Pb in the
nine NaI(Tl) crystals has been determined and preliminary results
are reported in the fifth and sixth columns of
table~\ref{detectors}. As made in previous set-ups, the potassium
content has been deduced by identifying coincidences between the
3.2~keV$_{ee}$ emissions and the 1460.8~keV gamma-ray following the
electron capture decay of $^{40}$K \cite{anaisijmpa}; the obtained
values are compatible with estimates from previous set-ups when
available. Some detectors have similar content to that of DAMA/LIBRA
crystals \cite{bernabei2008}; the average $^{40}$K activity in
ANAIS-112, although higher than that of DAMA/LIBRA, is more than one
order of magnitude lower than in large, low-background crystals
tested from other suppliers. The activity of $^{232}$Th and
$^{238}$U in the crystals is quantified by the measured alpha rates,
following Pulse Shape Analysis (to distinguish alpha interactions
from beta/gamma ones) and analysis of BiPo sequences; it is at a
level of a few $\mu$Bq/kg, but $^{210}$Pb out of equilibrium has
been observed for all the modules.
The origin of a possible $^{210}$Pb contamination was under study in
collaboration with AS, which allowed to obtain lower activity in the
last produced crystals (see table~\ref{detectors} and
figure~\ref{alphas}).

\begin{figure}
\begin{center}
\includegraphics[height=.37\textheight]{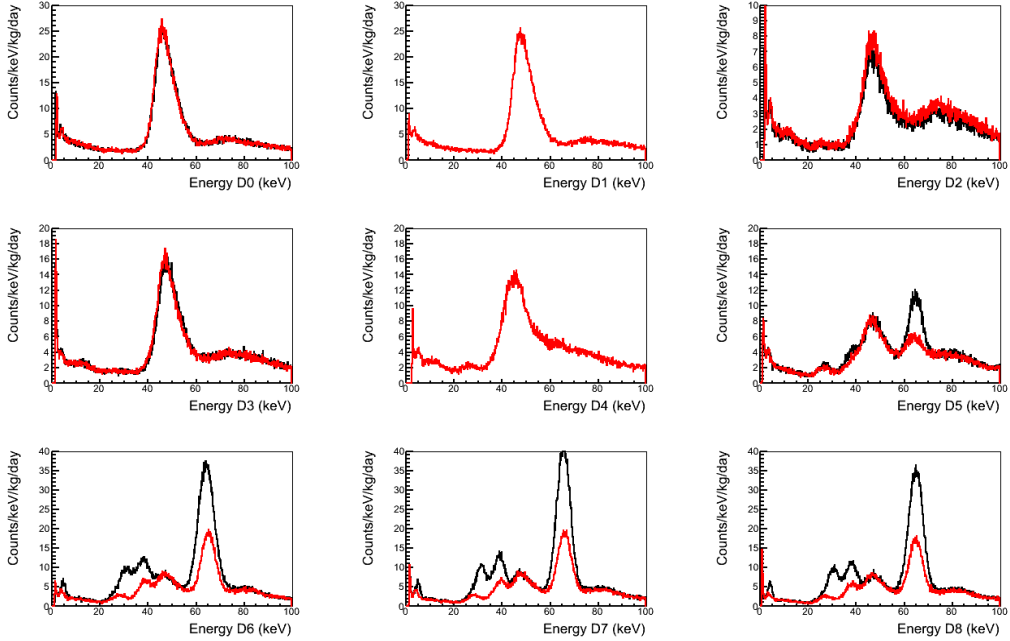} 
\end{center}
\caption{\label{bkgLE} Low energy region of the background spectra
registered in the commissioning run of ANAIS-112. Data from the
first 29.2~days (March and April, 2017) in black and the last
30.1~days (June and July, 2017) in red are shown (note they
correspond to filtered spectra but with no cut efficiency correction
yet).}
\end{figure}

\begin{figure}
\begin{minipage}{17pc} 
\includegraphics[width=\textwidth]{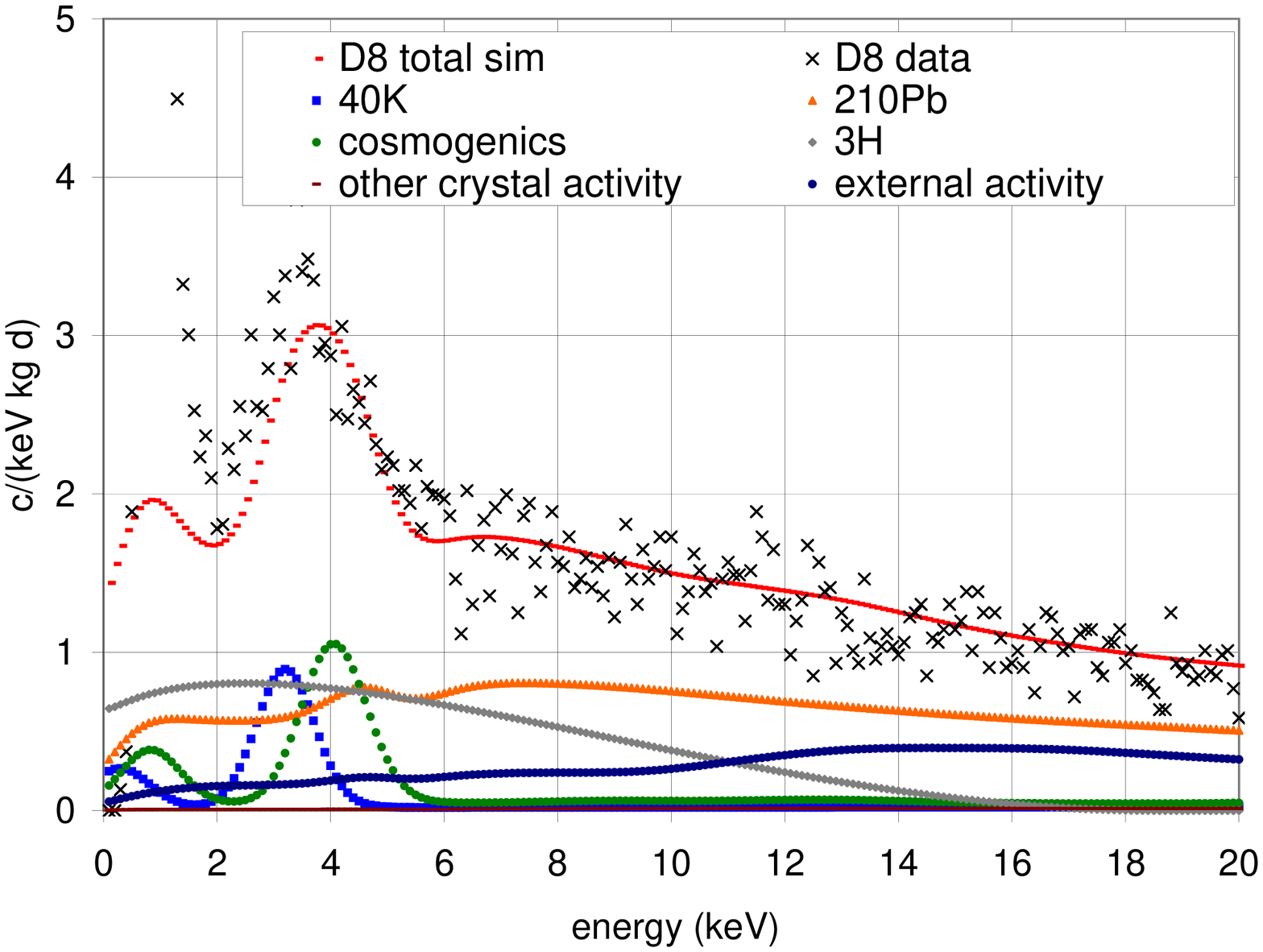}
\caption{\label{D8bkg} Comparison of low energy data taken in June
and July, 2017 with the background model from simulations for
detector D8. Individual contributions are shown too; cosmogenics is
still relevant due to the short time spent underground.}
\end{minipage}\hspace{1pc}
\begin{minipage}{20pc} 
\includegraphics[width=\textwidth]{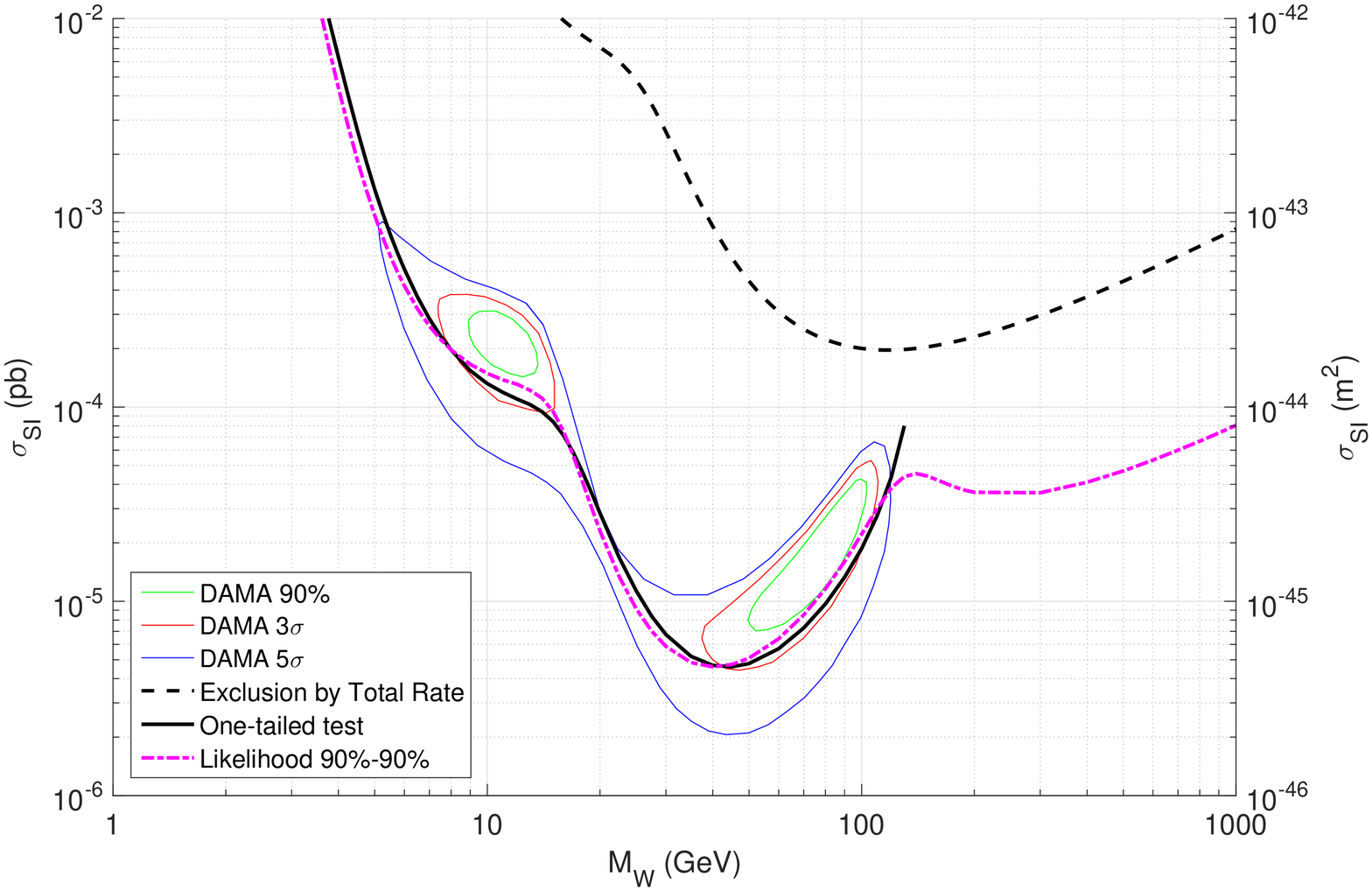}
\caption{\label{sensitivity} Annual modulation sensitivity prospects
for ANAIS-112 after 5 years of measurement, as evaluated in
\cite{anaisld}.}
\end{minipage}
\end{figure}

Figure~\ref{bkgLE} presents the background spectra at the low energy
region in ANAIS-112 commissioning run; first and latest data are
compared, showing the decay of cosmogenics in the last detectors.
Preliminary background models for those modules (see one example in
figure~\ref{D8bkg}), considering the measured crystal activities and
the ANAIS-112 configuration, point to equivalent relevant background
sources in the very low energy region. The $^{210}$Pb contribution
around 50~keV (see figure~\ref{bkgLE}) is consistent with the
measured alpha specific activity in all cases.

\section{Sensitivity prospects}
\label{sen}

The prospects of ANAIS-112 for the identification of an annual
modulation signal have been evaluated in \cite{anaisld} in terms of
the \emph{a priori} critical and detection limits of the experiment.
The analysis is based on the detector response and the background
level measured for the first modules operated in Canfranc. In
particular, an average background (corrected for the cut efficiency)
has been estimated in the regions of interest and five years of data
taking have been assumed. Considering the variance of the estimator
of the modulated amplitude, it is shown \cite{anaisld} that
ANAIS-112 in 2-6~keV$_{ee}$ has a detection limit for a
model-independent annual modulation (not related to a dark matter
origin) below the measured amplitude by DAMA/LIBRA
\cite{bernabei13}. As it can be seen in figure~\ref{sensitivity}
(taken from \cite{anaisld}), under the dark matter hypothesis, for a
detection limit at 90\% C.L. and a critical limit at 90\% C.L.,
ANAIS-112 can detect the annual modulation in the 3$\sigma$ region
compatible with the DAMA/LIBRA result.

\section*{Acknowledgments} A few days after TAUP2017,
Professor J.A.~Villar passed away. Deeply in sorrow, we all thank
his dedicated work and kindness. This work has been financially
supported by the Spanish Ministerio de Econom\'ia y Competitividad
and the European Regional Development Fund (MINECO-FEDER) under
grants No. FPA2011-23749 and FPA2014-55986-P, the Consolider-Ingenio
2010 Programme under grants MULTIDARK CSD2009-00064 and CPAN
CSD2007-00042 and the Gobierno de Arag\'on and the European Social
Fund (Group in Nuclear and Astroparticle Physics). We acknowledge
technical support from LSC and GIFNA staff.

\end{document}